# Artifact Reduction in Fundus Imaging using Cycle Consistent Adversarial Neural Networks

Sai Koushik S S, Dr. K.G. Srinivasa

*Abstract*—Fundus images are very useful in identifying various ophthalmic disorders. However, due to the presence of artifacts, the visibility of the retina is severely affected. This may result in misdiagnosis of the disorder which may lead to more complicated problems. Since deep learning is a powerful tool to extract patterns from data without much human intervention, they can be applied to image-to-image translation problems. An attempt has been made in this paper to automatically rectify such artifacts present in the images of the fundus. We use a CycleGAN based model which consists of residual blocks to reduce the artifacts in the images. Significant improvements are seen when compared to the existing techniques.

## I. Introduction

Fundus is an interior surface of the eye which is present opposite to the lens. It comprises the retina, optic disc, macula, fovea and the posterior pole [1]. Fundus images are two-dimensional images of the three-dimensional retinal tissues, captured using reflected light [2]. They are very useful in determining and monitoring the progression of certain eye diseases like glaucoma, diabetic retinopathy, abnormalities in the blood vessels, cotton wool spots, papilledema, multiple sclerosis and age-macular degeneration [3]. From an analysis performed on the retinal images of 5575 patients obtained from March to September 2003, it was found that 11.9% of the retinal images were of sub-standard quality [4].

Although this technique is a safe and non-invasive method to obtain images of the retina, numerous complexities are faced when reading these images. They mostly arise due to the presence of artifacts in the retinal images, which affect the visibility of the areas of interest in the retinal image. They can also lead to misdiagnosis of various optical conditions. For instance, false identification of the retinal nerve fiber layer (RNFL) can lead to misdiagnosis of significant glaucoma, which leads to undesired treatment [5]. Some artifacts in these images are shown in the figure below. (Fig. 1)

Image-to-image translation is a paradigm of deep learning applied to computer vision problems. The main objective of this is to understand the mapping between input and output images. It aims to transform an input image of certain undesired characteristics to an output image with the removal of the undesired elements, and may also involve the addition of certain desired characteristics. It is widely used in the fields of image processing, computer graphics and computer vision [6]. It is used for the purposes of image super-resolution [7], segmentation [8], style transfer [9], denoising [10] and colorization [11].

Deep learning-based techniques are very helpful in achieving this as they are very good at capturing complex patterns and can approximate mathematical functions with a very high accuracy. Neural networks are special mathematical models of the human brain [12] which are used to implement pattern recognition algorithms. The main objective of using these mathematical models is to capture these complex patterns and predict the outcome of a certain event. They are trained with pre-existing data in order to learn weights, which are used for the purpose of inference [13]. Their applications include, but are not restricted to function approximation [14], time series prediction [15], multivariate regression [16], pattern recognition and classification [17], biometric identification [18], data mining [19], spam filtering [20] and video upscaling [21].

Generative Adversarial Networks (GANs) are a class of deep learning techniques used for generative modelling. It is an unsupervised learning task which consists of two competing neural networks in a zero-sum game. The gain of one agent is considered as the loss of another during the process of training. The agents are referred to as *generator* and *discriminator* respectively [22]. For problems related to image generation the data from the training set, the objective is to synthesize new images while preserving the characteristics of the training set. They are widely used in the fields of reinforcement learning [23], semi [24] and fully supervised learning [25]. A schematic diagram of GANs is shown in the below figure. (Fig. 2)

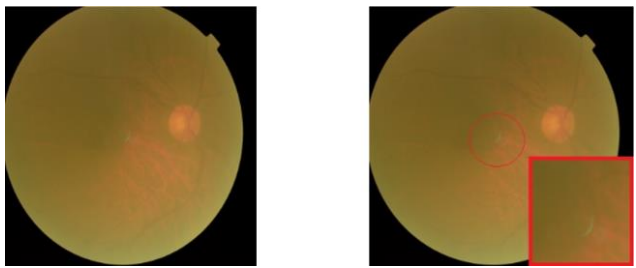

*Fig. 1*. *Left*: Image with artifact. *Right*: A closer view of the artifact. This image contains a lens flare, highlighted in the red region as shown. In addition to this, it also contains vignetted borders around the retina.

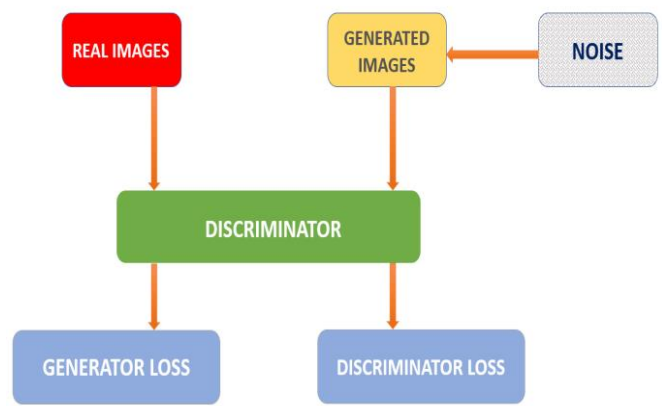

*Fig. 2.* Architecture of generative adversarial networks

GANs are very useful in image-to-image translation problems. Depending on the availability and nature of the dataset, suitable algorithms are chosen. CycleGAN [27] and pix2pix [26] are widely used techniques based on generative adversarial networks, and are used for image-to-image translation problems. While pix2pix requires paired images in the dataset to perform image-to-image translation, no such pairing in the training data is required to train CycleGAN models. This concept will be dealt with in Section III.

**Motivation for the work:** Since the idea of image-to-image translation can be utilized to rectify images containing artifacts with the resulting reconstruction being close to the ground truth, an algorithm based on cycle consistent generative adversarial networks is proposed. Significant improvements are seen when compared to the previous approaches. This can be helpful in reducing misdiagnosis due to wrong interpretations from the fundus images.

**Contribution of the paper:** In this paper, an attempt has been made to reduce the artifacts present in the fundus images of the retina using deep learning and generative modelling techniques. Improvements in the Perception based Image Quality Evaluator (PIQE) and Naturalness Image Quality Evaluator (NIQE) scores are seen.

**Organization of the paper:** This paper is organized into 8 sections. Sections I and II introduce the topic and mention the various approaches available in the existing literature. Section III mentions about the available techniques which can be applied to image-to-image translation tasks using deep learning. Section IV describes the dataset used in this paper. A description of the proposed solution and experimental results are explained in sections V and VI respectively. Section VII provides a comparative study between the proposed technique and available state-of-the art approach. Conclusion and future work is explained in Section VIII.

## II. RELATED WORK

Several authors have applied the concepts of image processing and computer vision to extract useful features from the fundus images. Zhang, Xiwei, et al. [28] proposed a technique to automatically identify normal images in tele-ophthalmology systems in order to reduce the time required for diagnosis of the images diagnosed with diabetic retinopathy. In addition to this, they also introduced a database which consists of manually contoured exudates. When compared to the already existing datasets, their dataset is very heterogeneous. They also proposed an algorithm to perform segmentation based on mathematical morphology by using new pre-processing techniques for denoising and normalization. On validating their algorithm on their own dataset, they obtained an Area Under Curve (AUC) of 0.95. Also, on validating their model on other datasets, they obtained AUCs in the range of 0.93 to 0.95. Martinez-Perez, et al. [29] proposed a technique to reconstruct retinal blood vessel trees using two views of fundus images. They used computer vision techniques which take into account various factors like the performing recovery of camera-eyeball model parameters using auto-calibration. This was done by utilizing simplified *Kruppa equations* which were based on correspondences found using a LMedS optimization correlation between pairs of eight views. They later attempted to extract skeletal and blood vessel information from these pairs of images. From the skeleton trees which were previously labelled, matching of corresponding points was made. Later, linear triangulation was performed on the correspondence points which were matched. This model was tested with two sets of 10 fundus images. The results of three-dimensional vein and artery reconstructions were also analyzed. Sierra, E., et al. [30] proposed an algorithm to identify and reduce the presence of dust particle artifacts from the images of the retina. Their method involved two steps. Initially, the retina image samples were checked for artifacts using normalized cross correlation with an artifact template, after which segmentation through region growing and comparison of the segmentations was performed in all the images, in order to ensure that detections are consistent for all the images being considered. The next stage was for removal which comprised of an inpainting algorithm which was done to suppress the visibility of the new region from that of the neighboring regions. Their results showed that such a technique was effective to remove artifacts and also ensured that no new artifacts were seen after the processing.

In addition to the traditional image processing techniques, solutions involving deep learning were also explored. Grinsven, Mark J. J. P. van, et al. [31] used convolutional neural networks to identify hemorrhage from fundus images. Since a lot of time is required to train CNNs, they proposed an algorithm to reduce the training time required for the purpose of training the model. It consisted of dynamically choosing wrongly classified false samples during the training process. They attempted to heuristically sample the data depending on the output status of the CNN. Later, weights were assigned to the samples. Samples which were informative were more likely to be selected for training the next iteration of the CNN. This approach allowed better accuracy along with the requirement of a smaller training time. They obtained an AOC value of 0.894 and 0.972 on their datasets. Yoo, Tae Keun, et al. [32] used CycleGANs to rectify the artifacts found in fundus images. They used a dataset consisting of 2206 anonymized images of the fundus. They resized all the images to a resolution of 256 × 256 before feeding them to the neural network. They used a simple CycleGAN architecture to rectify the artifacts in the fundus images. Their model architecture was the same as the one present in the tutorial for CycleGANs present in the documentation site for TensorFlow (Online: https://www.tensorflow.org/tutorials/generative/cyclegan)
They used a batch size of 1. For the optimization algorithm, they used the Adaptive Moment Estimation (ADAM) algorithm. On training the model for 40 epochs, they were able to reduce the artifacts from the fundus images. They demonstrated significant improvements from previously available approaches involving traditional image processing techniques. Sonali, et al. [33] implemented a technique to remove noise and enhace contrast in the images of the retina by integrating filters and contrast limited adaptive histogram equalization techniques. Their algorithm had an improvement of 7.85% in peak signal to noise ratio, 1.199% improvement in structural similarity index, 0.12% enhancement in correlation coefficient value when compared to the previously available methods.

On observing the related work, it is seen that not much work has been done towards automated enhancement of fundus images, specifically using deep learning techniques involving generative adversarial networks, with the exception of Yoo, Tae Keun, et al. [32]. Deep learning is a powerful approach to automate such image enhancement problems. Design of such automated systems will lead to the reduction in wrongly diagnosed cases due to misinterpretations of the retinal images.

In this paper, an improvised technique to reduce artifacts in fundus images is proposed. Significant improvements are seen when compared to the work presented by Yoo, Tae Keun, et al.

### III. AVAILABLE APPROACHES

As mentioned earlier (Section I), there are two widely used techniques to solve image-to-image translation problems: pix2pix [26] and CycleGAN [27]. Although both algorithms are used widely, the main limitation of pix2pix is the requirement of paired datasets. That is, every image in one domain should have its equivalent in the other domain in the dataset used for training. There are two neural networks in this algorithm: Generator and Discriminator. A random image with noise is provided as input to the generator neural network. This network then performs mathematical operations called as *transpose convolutions* on this noise in multiple layers. The final output layer then produces images which are intended to be in the desired category or domain. In this model, the output of the generator is conditional on the source image. The discriminator neural network has two inputs: source image and target image. The output of the discriminator is used to determine the feasibility of the transformation in the domain of the source image. We then proceed to train the generator neural network using adversarial loss. The gain of one network is considered as the loss of the another. This concept is applicable to the generator-discriminator arrangement. The L1 loss is then calculated by considering the generated output and the expected output. In case the loss function of the generator increases, then this is penalized by decreasing the loss function of the generator as the main objective of this approach is to minimize the loss function of the generator and maximize the loss function of the discriminator. On training the networks for several approaches, the generator network then produces images with a realistic appearance. Suitable choice of optimizer function also plays a major role in the training process.

For image-to-image translation problems, pix2pix is used for converting maps to satellite images, facades to real buildings, single channel images to three channel images (black and white to color images), etc.

CycleGAN [27] is another approach to this problem. It is possible to extract useful information from the training dataset while not requiring pairings in the set of images used for training purposes. If the translation problem involves two domains, $M$ and $N$. Our objective is the translation of domains, $M \rightarrow N$.

To obtain such a translation, two generator models are involved. The first generator neural network ($G_M$) aims to generate images in the first domain, $M$. Similarly, the second generator neural network ($G_N$) is used to produce images in the second domain, $N$. The constraints are as follows:

a) The input to $G_M$ is an image from domain $N$.
b) The input to $G_N$ is an image from domain $M$.

In other words, the generator, $G_M$ takes an image from domain $N$ and tries to convert it to domain $M$ by performing successive *transpose convolutions* in successive layers. Similarly, $G_N$ takes an image from domain $M$ and tries to convert it to domain $N$.

Since it is very difficult to obtain and annotate thousands of paired images, it is inconvenient to use pix2pix based approaches. Hence, models based on unsupervised CycleGANs are more robust and versatile when compared to pix2pix. Due to this reason, CycleGAN based approach was followed for this translation task.

### IV. DATASET DESCRIPTION

The proposed method involves collecting images belonging to two categories: Images of the funds with the presence of artifacts and images with the absence of artifacts. On surveying the available information, it was found that using the data contained in the dataset consists of 2206 images, categorized into two categories.

All the data contained in the dataset is completely anonymized and available on Mendeley Data platform by TaeKeun Yoo [34] . Out of the 2206 images, 1146 images contained artifacts and 1060 images did not have any artifacts. The experimentation process involved in gathering of the data was done in conformation to the Declaration of Helsinki. Some images from this dataset are shown below in *Fig. 3* and *Fig. 4*.

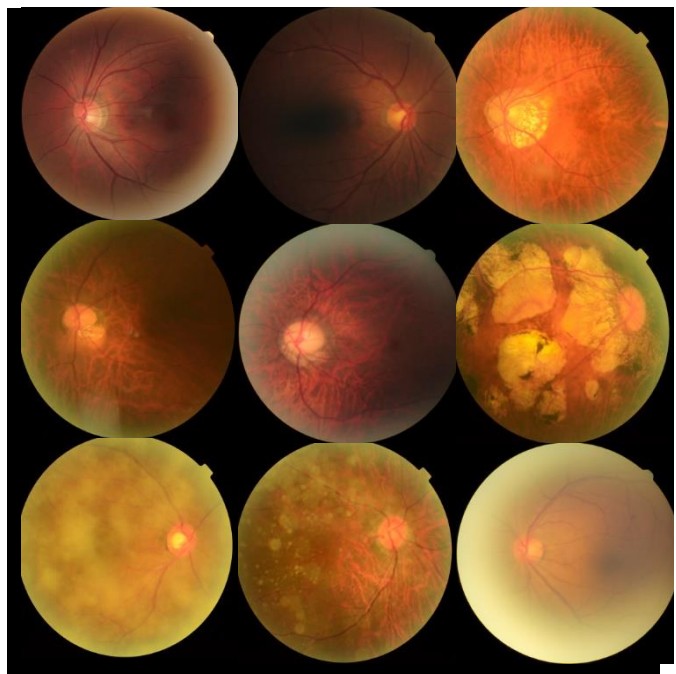

*Fig. 3: Some images of the fundus from the dataset containing artifacts*

In the above figure, (Fig. 3) some images containing artifacts are shown. These were randomly selected from the dataset. Arc artifacts, haze artifacts, artifacts arising due to improper illumination, images with lens flare, lash artifacts are some errors found in the dataset images, in general.

The visibility of the retina is severely affected due to such unwanted elements in the dataset, which we attempt to eliminate in this study.

As per the notations used in this paper, the images of this category belong to domain *M*.

Sufficient care was taken to ensure that by using the CycleGAN based approach for image-to-image translation, the basic structure of the image was not affected in any way.

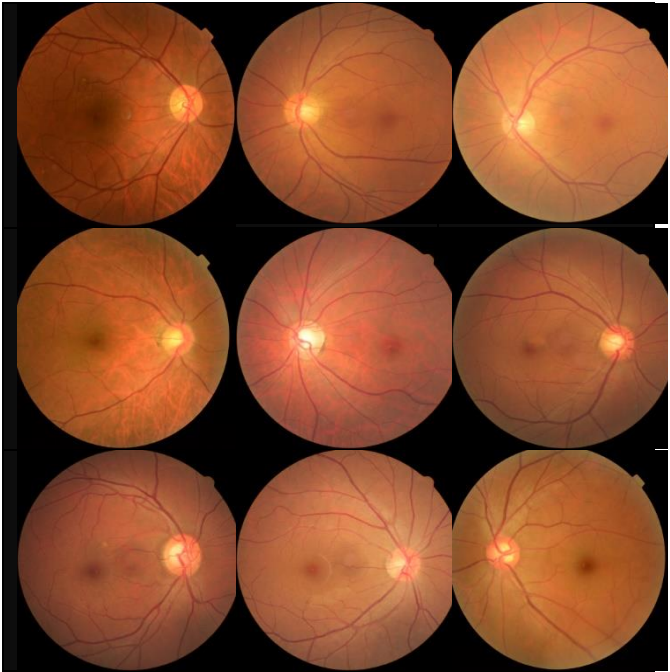

*Fig. 4: Fundus images without any artifacts*

The images shown in Fig.4 belong to domain *N*. These images taken from the dataset do not contain any artifacts. This is the target domain.

## V. PROPOSED TECHNIQUE

The input to the first discriminator neural network, $D_M$ is the image generated by $G_M$. It also takes images from domain *M* as input and predicts the *trueness* of the generated image. In other words, the output of this discriminator is a measure of the *closeness* of the generated image to the domain *M*. Similarly, the discriminator network $D_N$ takes the image generated by $G_N$ in addition to the actual images from the dataset belonging to the domain *N*. The output of $D_N$ is a measure of the *closeness* of the generated image to the domain *N*. Now, the generator and discriminator neural networks are trained in an adversarial manner. Furthermore, an additional *identity mapping* is introduced to this architecture. This is done in order to enhance the matching of color profiles of input and target domain images. We provide the generator with images from the target domain and expect it to generate a replica of the exact image without any changes. Although this is optional for the implementation of CycleGANs, this was useful to enhance the robustness of the model.

Before defining the model architecture, some pre-processing of the images was required. The images were resized to 256×256 dimensions.

There are four neural networks present in CycleGAN technique: Two generators and two discriminators.

A multilayered deep convolutional neural network architecture was used for the discriminator networks. Three important types of layers are present in this model architecture: Convolutional Layers, Instance Normalization and Activation Layers.

**Need for normalization for this specific problem:** The idea of normalization is very important and is an area of research in the field of deep learning [35] [36] [37]. The utilization of these techniques can significantly reduce the time required to train the neural network. The main advantages of normalization technique are:

i) The contribution of all features is considered for the output or prediction of the neural network. This means that the dominance of certain features with greater magnitude over the features with less significant magnitude is reduced, thus reducing the notion of *bias* in the neural network architecture.

ii) Reduction in internal covariate shift. The changes in the distribution of neural network activation functions vary with the change in the parameters of the neural network when training takes place. It is necessary to ensure that such changes in the distribution do not happen. Normalization is helpful in achieving this objective.

iii) Speed up the process of optimization as weights are not changing sharply. Their values are restricted to specific ranges due to their magnitudes being restricted. It is also due to the fact that the gradient magnitudes are bound tightly by smoothening the loss surface. [38]

The most widely used normalization technique is referred to as *Batch Normalization* [39]. In this approach, the activation functions in a network are normalized over a mini batch of pre-determined size. The mean and variance of the features is calculated for every batch. For all the features, the individual magnitude is subtracted with the mean and the resulting value is divided with the mini-batch standard deviation. Mathematically, we denote this as:

$$\mu_f \leftarrow \frac{1}{n}\sum_{i=1}^{n} x_i \qquad (1)$$

$$\sigma_f^2 \leftarrow \frac{1}{n}\sum_{i=1}^{n}(x_i - \mu_f)^2 \qquad (2)$$

Here, the quantity $\mu_f$ is the mean for the specific mini-batch. $\sigma_f^2$ is the variance computed for every mini-batch.

In addition to this, it is also important to adjust the weight magnitudes to ensure that the neural network performance is enhanced. This is done by introducing quantities $\hat{x}_i$ and $y_i$ to account for normalization and scaling respectively.

$$\hat{x}_i \leftarrow \frac{x_i - \mu_f}{\sqrt{\sigma_f^2 + \delta}} \qquad (3)$$

$$y_i \leftarrow \gamma \hat{x}_i + \beta \qquad (4)$$

The quantity β is a constant, and is used to enhance the stability during the process of training the neural network.
The main disadvantage of using this technique is that the model output becomes very noisy due to which training is affected when the mini-batch size is small. A smaller batch size was needed for this problem considering the availability of computational resources. Hence, *instance normalization* [40] technique was used. This algorithm attempts to normalize across every channel in every individual training input rather than attempting to normalize across features in the input. It can also be applied during testing of the model as it does not depend on the mini-batch. This concept is extensively used in GANs.

In this algorithm, the mean specific to the instance and also the shift in covariance is suppressed, leading to a simplified model learning. It attempts to reduce contrast information specific to the instance from the training example so that the process of generating new samples is simplified [40] [41].

$$\mu_f = \frac{1}{pq} \sum_{i=1}^{p} \sum_{j=1}^{q} x_{fij} \qquad (5)$$

$$\sigma_f^2 = \frac{1}{pq} \sum_{i=1}^{p} \sum_{j}^{q} (x_{fij} - \mu_f)^2 \qquad (6)$$

$$\hat{x} = \frac{x - \mu_f}{\sqrt{\sigma_f^2 + \delta}} \qquad (7)$$

In this approach, we calculate the mean ($\mu_f$) and variance ($\sigma_f^2$) for every sample across all spatial dimensions. The output values of every feature map are scaled to a standard Gaussian distribution here.

For each discriminator network, the input is of $256 \times 256$ dimensions. It consists of 5 convolutional layers comprising of 64, 128, 256, 512 and 512 units in every successive layer respectively. The stride size was set to (2,2). For every layer, the LeakyReLU function [42] is used as activation. This function is mathematically denoted as:

$$f(x) = \begin{cases} x, & \forall x \geq 0 \\ \alpha x, & \forall x < 0 \end{cases} \qquad (8)$$

The quantity α is the slope value which is adjusted as necessary [43]. The value of $\alpha = 0.325$ was used in this problem. These values were chosen experimentally.

The last layer of the discriminator network consists of a single unit. The model was compiled using the Adaptive Moment Estimation (Adam) algorithm [44] with a learning rate of 0.000364 and β value of 0.5032. Other optimizer functions like Stochastic Gradient Descent [46] (SGD) could have been used, but it has certain limitations. SGD introduces stochasticity to the gradient descent algorithm whereas Adam incorporates adaptive gradients in which different learning rates are applied for different parameters. It also computes and accounts for the exponentially decaying average of past squared gradients, which adds *momentum* to the optimization process. The update process of this function is shown in the equations that follow this section. It can be visualized as the addition of momentum to the stochastic gradient descent optimizer, which results in this function.

$$v_x = \beta_1 v_{(x-1)} + (1 - \beta_1) q_x \qquad (9)$$

$$s_x = \beta_2 l_{(x-1)} + (1 - \beta_2) q_t^2 \qquad (10)$$

In the above equations, $v_x$ is the first moment which accounts for the momentum occurring at the past normalized gradients. $s_x$ is the second moment estimate, that is used for assigning different learning rates for multiple parameters.

During the process of training the neural network, the initial values of these parameters of moments are equal to zero. Due to this, they are biased towards zero in the initial training steps, when decay rates are small. Hence, we utilize bias corrected moment estimates, given by the equations that follow.

$$v'_x = \frac{v_x}{1 - \beta_1^x} \qquad (11)$$

$$s'_t = \frac{s_x}{1 - \beta_2^x} \qquad (12)$$

Thus, incorporating these results in the equations 9 and 10, the resulting update rules are obtained [44] [46]:

$$g'_x = \frac{(learning\ rate) \times v'_x}{\sqrt{s'_t + \delta}} \qquad (13)$$

$$\rho_x = \rho_{x-1} - g'_x \qquad (14)$$

For our problem, we used a learning rate of 0.000364. This value was also determined experimentally. For the loss function, least squares loss was used. This function is given by:

$$L = \sum_{i=1}^{r} (y_{actual} - y_{computed})^2 \qquad (15)$$

For the generator neural networks, a combination of encoder and decoder architecture is used. The input to the generator is the image from the source domain whereas the output is an image from the target domain. The source domain image is taken and down sampled to a certain layer. This is done to perform encoding. After this, the encoding is fed to Residual Network layers (ResNet) which involve skip connections. They are modelled on the pyramidal cells present in the cerebral cortex. This functionality is approximated by utilizing skip connections to jump over certain layers. There may be two or three layer skipping which consist of activations and batch normalizations [46] [47].

This is done in order to simplify the neural network so that a smaller number of layers are used in the initial phases due to which learning is accelerated as the number of layers for propagation is less. As the feature space is gradually learnt during training, the skipped layers are gradually restored. Thus, the exploration of feature space [48] [49] is restricted for the neural network involving residual layers so that unwanted disturbances during training are prevented.

A canonical form of these networks, is shown in the below figure.

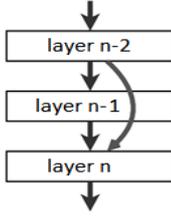

*Fig. 5. Canonical form of ResNets*

In the above figure, it can be noted that layer *n-1* is skipped over the output activation of layer *n-2*.

The skip connections serve two purposes. They help to reduce the *vanishing gradients* problem [53]. This occurs when the weights do not update during the training process. Since residual networks use skip connections to permit the passage of gradient information through the layers, it is possible for the information transfer between the earlier layers and the deeper layers of the neural network, so that signal propagation is enhanced even if the architecture of the neural network is deep [46].

In a residual network, if the weight matrix is $W^{n-1,n}$ denotes the connecting weights from the layer *n-1* to layer *n* and if $W^{n-2,n}$ represents the connecting weights from layer *n-2* to *n*, the output of the neurons present in the *n*th layer is expressed as:

$$A^n := f(W^{n-1,n} \cdot A^{n-1} + B^n + W^{n-2,n} \cdot A^{n-2}) \quad (16)$$

or,

$$A^n := f(Q^n + W^{n-2,n} \cdot A^{n-2}) \quad (17)$$

In the above equations, the quantity *f* is the activation of neurons which are present in the *n*th layer and

$$Q^n = W^{n-1,n} \cdot A^{n-1} + B^n \quad (18)$$

If the skipping occurs for several layers, equation 17 can be expressed as:

$$A^n := f\left(Q^n + \sum_{v=2}^{V} W^{n-v,n} \cdot A^{n-v}\right) \quad (19)$$

Similarly, the backpropagation algorithm is modified. The learning of parameters in regular paths in the neural network during back propagation can be expressed as:

$$\Delta W^{n-1,n} := -\alpha \frac{\partial H^n}{\partial W^{n-1,n}} = -\alpha A^{n-1} \cdot \tau^n \quad (20)$$

For the paths involving skip connections, this equation is modified as,

$$\Delta W^{n-2,n} := -\alpha \frac{\partial H^n}{\partial W^{n-2,n}} = -\alpha A^{n-2} \cdot \tau^n \quad (21)$$

If skipping occurs for multiple layers, the equation above may be generalized as:

$$\Delta W^{n-v,n} := -\alpha \frac{\partial H^n}{\partial W^{n-v,n}} = -\alpha A^{n-v} \cdot \tau^n \quad (22)$$

$\alpha$ is the learning rate, $\tau^n$ indicates the error signal at the *n*th layer.

For this implementation, every ResNet block in the generator consists of two convolutional layers, with each layer being followed by instance normalization. The filters are of dimensions (3×3) and stride is of (1×1) size.

The generator architecture consists of three convolutional layers with 64, 128 and 256 filters in each successive layer respectively. 9 blocks of ResNets follow the third convolutional layer. The residual blocks are then followed by two successive transpose convolutional layers. Here also, every transpose convolutional layer is followed by instance normalization operation. The two transpose convolutional layers consist of 128 and 64 units (filters) respectively. The instance normalization of the second transpose convolutional layer is followed by the *tanh* activation function.

The input to the discriminator network is the data corresponding to both real and synthesized images. The block diagrams in Fig. 6 show the architecture of the generator model and the discriminator model.

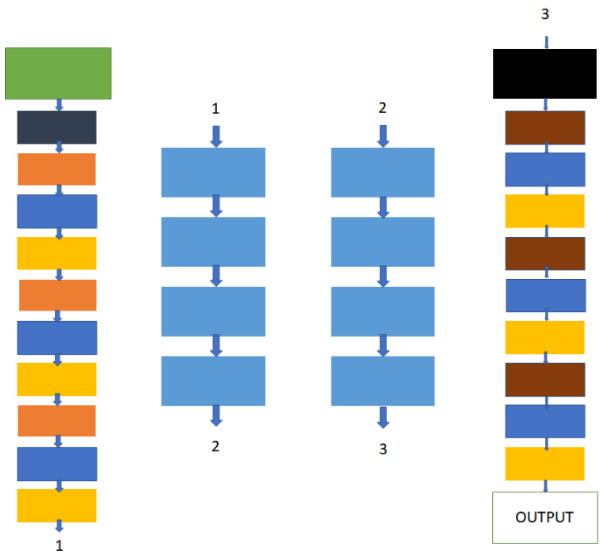

*Fig. 6(a): Architecture of the generator model: Block diagram*

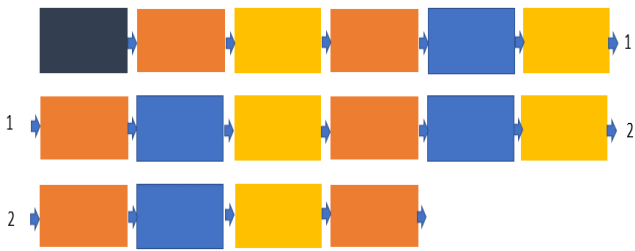

*Fig. 6(b): Architecture of the discriminator model*

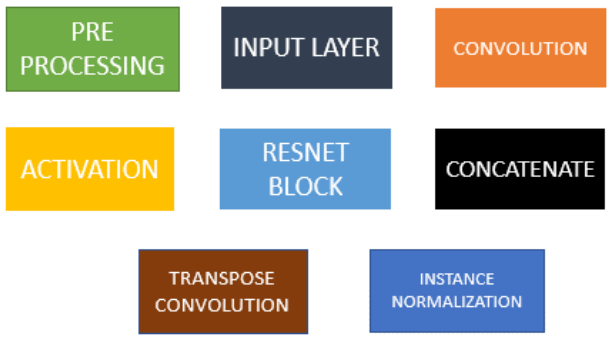

*Fig. 6(c): Legend for the neural networks depicted in figures 6(a) and 6(b)*

The model was trained for 200 epochs on the Google Colab online platform on a Linux virtual machine with 16 Giga Bytes of RAM, Intel Xeon CPU running at 2.20 GHz, Tesla T4 graphics processing unit with a Nvidia TU104 graphics processor, based on the Turing GPU architecture. The GPU consists of 16 Giga Bytes of dedicated graphics memory to speed up the model training process.

## VI. EXPERIMENTAL RESULTS

On training the model for 200 epochs, realistic results were obtained. Some images of the output are shown below.

The input image to the CycleGAN is shown in the left and the corresponding output for the specific image is shown in the right.

| | INPUT IMAGE | RECTIFIED IMAGE |
|---|---|---|
| A | | |
| B | | |
| C | | |
| D | | |
| E | | |
| F | | |

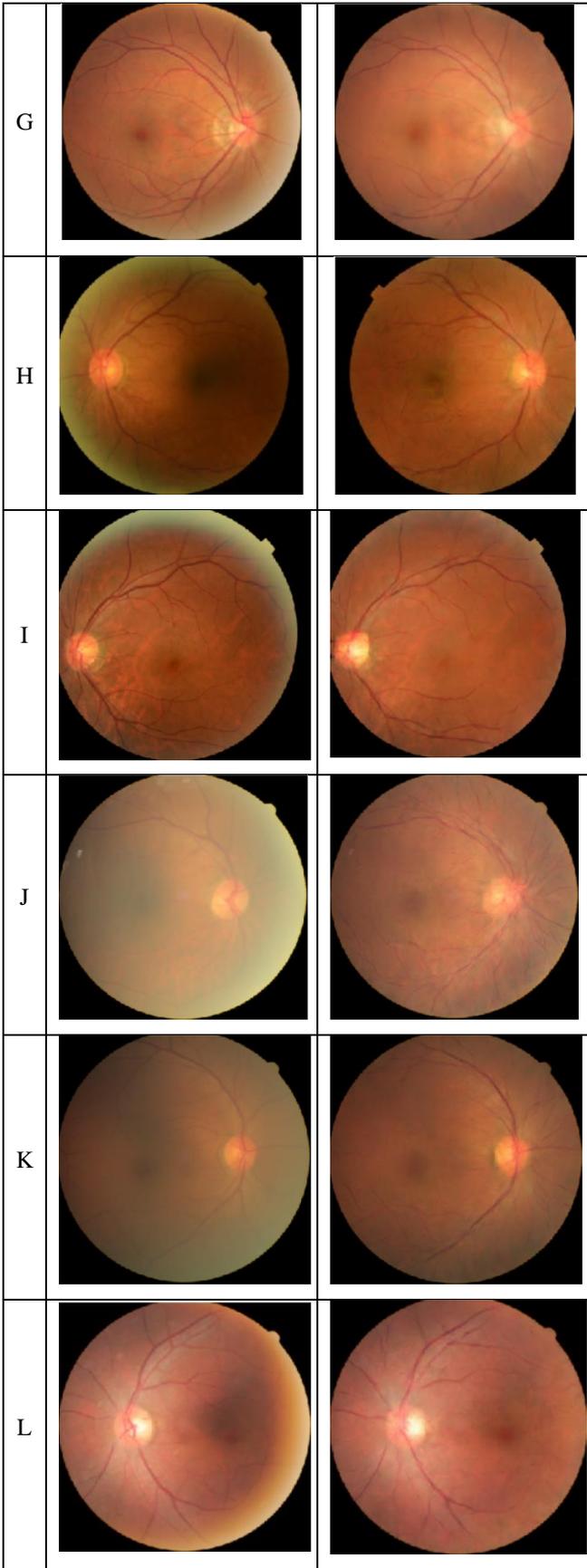

*Fig. 7. Some sample images taken from the testing set. The images on the left are the input images. The corresponding output for every image is shown in the right.*

Some images from the test set are shown in Fig. 7. Arc artifacts (B,C,H,I), haze artifacts (D,J,K), artifacts arising from improper illumination (L, F) and opaque artifacts (A,E) are shown along with their rectified images. It can be observed that the fundamental structure of the image is not altered. Also, new artifacts are not being introduced in the fundus image, when using this approach.

## VII. DISCUSSION

The results obtained are compared with that of (Yoo, Tae Keun, et al.) [32]. A detailed comparison of the same is made in this section.

In the work done by Yoo, Tae Keun, et al. [32] the generator consisted of a U-Net based model architecture for the generator models, which was taken directly from the documentation webpage for CycleGAN available on the TensorFlow website.

U-Nets consist of convolutional layers, max-pooling layers, ReLU activation, concatenation, and transpose convolutional layers. There are three main blocks in all U-Nets: Contraction, bottleneck and expansion blocks.

In contrast to this, the approach described in this paper uses the Leaky ReLU activation function (Section V, Equation #8). Leaky ReLU assigns a smaller magnitude of slope for negative values which may be present at the output of the units present in every layer of the neural network. In the case of ReLU activation function, the output is completely assigned to a magnitude of 0 whenever there is a negative input to the activation function because for a ReLU activation function,

$$f(x) = \begin{cases} x, & \forall\, x \geq 0 \\ 0, & \forall x < 0 \end{cases} \quad (16)$$

Thus, our approach considers the effect of the magnitudes of the convolutional outputs even if they are less than 0 rather than completely neglecting them, which further enhances the robustness of the model by capturing better patterns from these ignored values.

Residual Networks (ResNets) require a lesser number of transpose convolutions in the neural network architecture in comparison with U-Nets. U-Nets require several transpose convolutional layers for the same. The additional transpose convolutional layers require more computational resources and as a consequence, a greater amount of training time is required. These may be eliminated by the inclusion of ResNets in the model architecture.

The generator network used by [32] utilizes batch normalization as opposed to instance normalization, which is used in this paper. As discussed in *Section V*, instance normalization computes mean and variance for every channel individually rather than an entire mini-batch. Hence, instance normalization makes the image distribution more Gaussian to the neural network during the training process. In addition to this, unwanted noise is added to the data while training the neural network when batch normalization is used. Thus, this effect of noise negatively affects the performance of the neural network while generating new images. Thus, instance normalization is more applicable for image-to-image translation tasks. Also, the contrast variations in the input images do not affect the model during the process of training.

For a detailed comparative study, three major artifacts were not rectified in the technique implemented by [32]. The model proposed in this paper was tested for such errors and the output for the same is shown in Fig. 8. It may be noted that a lot of artifacts, if not all were rectified in the output image, with better preservation of details than that of Yoo, Tae Keun, et al. Asteroid Hyalosis based artifacts, severe uneven illumination artifacts and severe haze artifacts are analyzed and shown in the above figure. It may be noted that such imperfections can be better enhanced by using this CycleGAN architecture.

| Some improvements in the proposed method for correcting fundus images ||
|---|---|
| Yoo, Tae Keun, et al. [32] | Proposed Technique |
| **Left**: Input Image. **Right**: Rectified Image | **Left**: Input Image. **Right**: Rectified Image |
| **Asteroid Hyalosis** ||
| **Severe Uneven Illumination Artifacts** ||
| **Severe Haze Artifacts** ||

*Fig. 8. Comparison of the two models*

It is also worth mentioning that the model performs well when provided with images with severe glaze artifacts, when compared to [32].

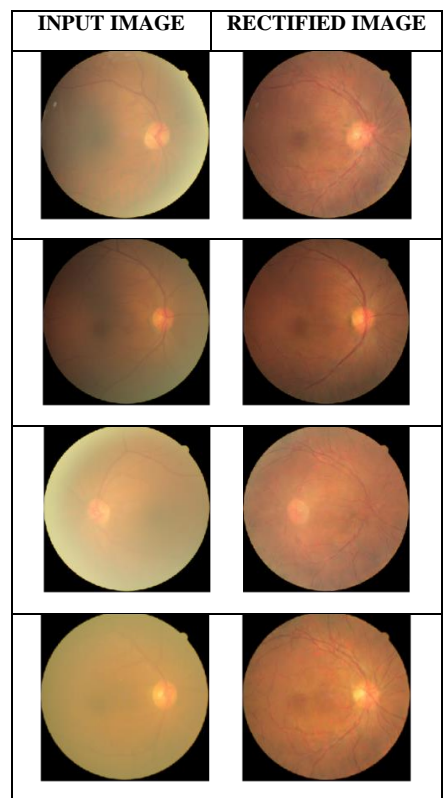

*Fig. 9. Severe glaze artifact rectification using this technique*

To test the model from a mathematical perspective, Perception based Image Quality Evaluator (PIQE) and Naturalness Image Quality Evaluator (NIQE) scores are calculated for the images from the testing set. 352 images were used for testing the model. A plot of the NIQE scores is shown in the below (Fig. 10) figure. This score is obtained by calculating the distance between the natural scene statistical features, which are modeled as Gaussian distributions with multiple dimensions from an input image and the features derived from an image dataset while training the model. A custom model for this assessment was trained using the images without any using the fundus images dataset used in this article. On obtaining the custom model in MATLAB, the output images of the CycleGAN were compared against the base NIQE model which is taken as reference. A plot of these values for all the 352 images is shown in the figure below. It is to be noted that a lower NIQE score indicates more perceptual quality. These scores are computed for the input images to the CycleGAN (shown in blue) and the output images from the CycleGAN (shown in red). It can be seen that the score for the output images is lesser than that of the input images.

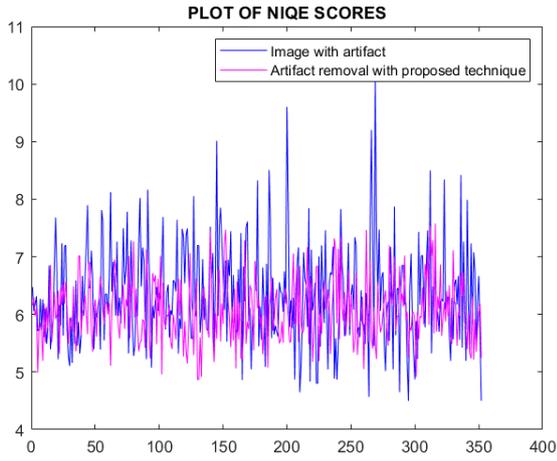

*Fig. 10. A plot of NIQE scores for the images in the testing set.*

Similarly, a plot of the PIQE scores is shown in Fig. 11. This score is also inversely proportional to the perceptual quality of an image. It is a positive number, in the range of 0 to 100. This score is computed using the following process using a technique known as block-wise estimation of distortions [50] [51]:

(1) Calculation of the Mean Subtracted Contrast Normalization (MSCN) coefficients for every individual pixel present in the luminance image $I$ using the equations proposed by Mittal, Anish, et al. [52]. These equations are:

$$\hat{I}(x,y) = \frac{I(x,y) - \mu(x,y)}{\sigma(x,y) + \varepsilon} \quad (17)$$

Here, $x = \{1,2,3, \ldots, X\}$ and $y = \{1,2,3, \ldots, Y\}$

$x$ and $y$ are referred to as the *spatial indices*. $X$ and $Y$ are the dimensions of the image. The quantity $\varepsilon$ is used as a stabilizing constant in the above equation. Mittal, Anish, et al. [52] also proposed the process to obtain these values using the following expressions:

$$\mu(x,y) = \sum_{g=-G}^{G} \sum_{h=-H}^{H} f_{g,h} Z_{g,h}(x,y) \quad (18)$$

$$\sigma(x,y) = \sqrt{\sum_{g=-G}^{G} \sum_{h=-H}^{H} f_{g,h} (Z_{g,h}(x,y) - \mu(x,y))^2} \quad (19)$$

In the above equations, $f = \{f_{g,h} | g = -G \ldots, G, h = -H, \ldots, H\}$

It is a Gaussian filter which is circularly symmetric. The MSCN coefficients are thus obtained from the above equation (Equation 17).

(2) The image is then divided into $16 \times 16$ blocks with no overlapping.

(3) Blocks which are spatially active are determined using the variance of the MSCN coefficients, as described in the equations 17 through 19.

(4) Evaluation of distortion in individual blocks. These distortions result from artifacts and image noise.

(5) Using the threshold criteria proposed by [50], blocks are classified as distorted or undistorted blocks. If there is distortion, the blocks are further classified as distorted blocks with blocking artifacts or blocks with Gaussian noise.

(6) PIQE score is then given as the average of the scores computed for every individual distorted block.

A plot of this score (lower is better) is shown in the below figure. (Fig. 11) Similar to the calculation of NIQE scores, these scores are computed for the input images for the CycleGAN (shown in blue) and the output images from the CycleGAN (shown in red). Significant reduction at the output can be noticed in the graph.

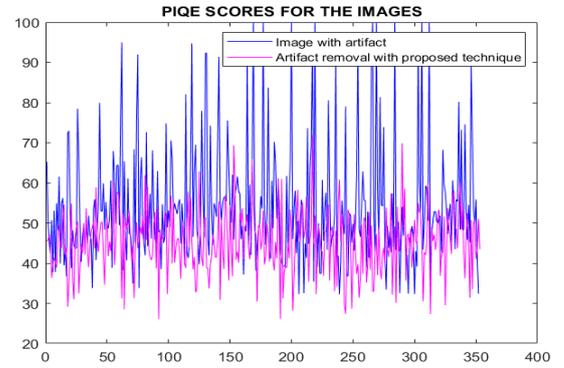

*Fig. 11. A plot of PIQE scores for the images in the testing set*

A graph of the NIQE and PIQE scores of the proposed technique and that of Yoo, Tae Keun, et al. [32] is shown in the graphs below (Fig 12 and Fig 13).

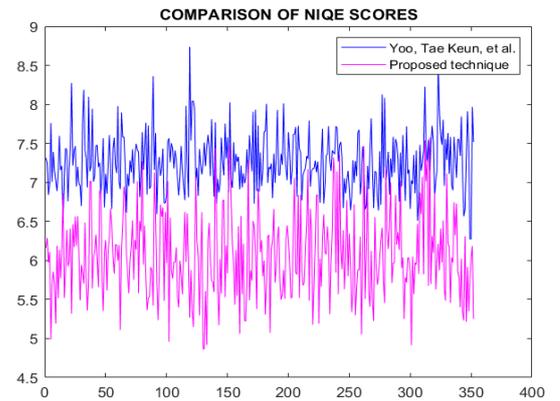

*Fig. 12. Comparison of NIQE scores*

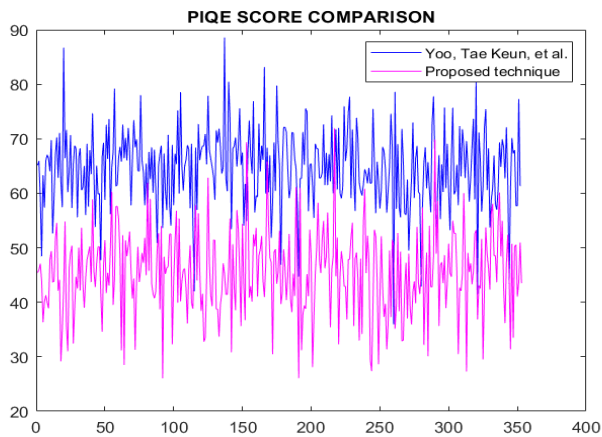

*Fig. 13. Comparison of PIQE scores*

## VIII. CONCLUSION

In this paper, an attempt has been made to rectify the artifacts in the fundus images. Due to various factors, errors are introduced in the fundus image which affect the visibility of various parts of the retina in the image. These may lead to false diagnosis of various diseases, as described in *Section I*. Many authors mentioned in *Section II* used various image processing techniques in rectifying the artifacts (arc artifacts, severe glaze artifacts, artifacts resulting from improper illumination, asteroid hyalosis, and lashes-based artifacts) present in the fundus images. However, deep learning techniques act as powerful approximators of highly complicated mathematical functions and thus can be applied to pattern recognition problems, due to which we utilized a CycleGAN based model for this problem. Improvements to the existing approaches are also shown.

In future, this may be extended to better accuracy levels by using more powerful high performance computing hardware. The rectification of artifacts by the models can further be enhanced by harnessing the power of transfer learning. More complex datasets may also be used to obtain more rich, powerful and efficient models for such image-to-image translation problems. This same architecture can also be extended for other medical problems involving translation like translating Computerized Tomography (C.T) images to Magnetic Resonance Images (M.R.I), X-Ray to CT translation etc.

## REFERENCES

[1] Cassin, B. and Solomon, S. Dictionary of Eye Terminology. Gainesville, Florida: Triad Publishing Company, 1990.

[2] Amit BJ, Jaya Prakash V, Muna B. Techniques of Fundus Imaging, Sci J Med & Vis Res Foun 2015;XXXIII:100–107.

[3] M. D. Abramoff, M. K. Garvin, and M. Sonka, "Retinal Imaging and Image Analysis," IEEE Reviews in Biomedical Engineering, vol. 3, pp. 169–208, 2010

[4] Philip S, Cowie LM, Olson JA. The impact of the Health Technology Board for Scotland's grading model on referrals to ophthalmology services. Br J Ophthalmol. 2005 Jul;89(7):891-6. doi: 10.1136/bjo.2004.051334. PMID: 15965173; PMCID: PMC1772738.

[5] Stuart, Annie. "What's Wrong With This Picture?" American Academy of Ophthalmology, American Academy of Opthalmology, 28 Apr. 2016, www.aao.org/eyenet/article/whats-wrong-with-this-picture.

[6] Sung TL, Lee HJ. Image-to-Image Translation Using Identical-Pair Adversarial Networks. Applied Sciences. 2019; 9(13):2668. https://doi.org/10.3390/app9132668

[7] Nasrollahi, K.; Moeslund, T.B. Applications. Super-resolution: A comprehensive survey. Mach. Vis. 2014, 25, 1423–1468. https://dx.doi.org/10.1007/s00138-014-0623-4

[8] Khan, M.W. A survey: Image segmentation techniques. Int. J. Future Comput. 2014, 3, 89. https://dx.doi.org/10.7763/IJFCC.2014.V3.274

[9] Gatys, L., Ecker, A., & Bethge, M. (2016). A neural algorithm of artistic style. Journal of Vision, 16(12), 326. doi:10.1167/16.12.326

[10] Elad, M.; Aharon, M. Image denoising via sparse and redundant representations over learned dictionaries. IEEE Trans. Image Process. 2006, 15, 3736–3745. https://dx.doi.org/10.1109/TIP.2006.881969

[11] Luan, Q.; Wen, F.; Cohen-Or, D.; Liang, L.; Xu, Y.-Q.; Shum, H.-Y. Natural image colorization. In Proceedings of the 18th Eurographics Conference on Rendering Techniques, Grenoble, France, 25–27 June 2007; pp. 309–320.

[12] Hopfield, John J. "Neural networks and physical systems with emergent collective computational abilities." Proceedings of the national academy of sciences 79.8 (1982): 2554-2558.

[13] N.L.W. Keijsers, Neural Networks, Editor(s): Katie Kompoliti, Leo Verhagen Metman, Encyclopedia of Movement Disorders, Academic Press,2010, Pages 257-259, ISBN 9780123741059, https://doi.org/10.1016/B978-0-12-374105-9.00493-7. (https://www.sciencedirect.com/science/article/pii/B9780123741059004937)

[14] S. Ferrari and R. F. Stengel, "Smooth function approximation using neural networks," in IEEE Transactions on Neural Networks, vol. 16, no. 1, pp. 24-38, Jan. 2005. doi: 10.1109/TNN.2004.836233

[15] Zhang, G. Peter. "Time series forecasting using a hybrid ARIMA and neural network model." Neurocomputing 50 (2003): 159-175.

[16] Cui, Chenhao, and Tom Fearn. "Modern practical convolutional neural networks for multivariate regression: Applications to NIR calibration." Chemometrics and Intelligent Laboratory Systems 182 (2018): 9-20.

[17] Alex Krizhevsky, Ilya Sutskever, and Geoffrey E. Hinton. 2017. ImageNet classification with deep convolutional neural networks. Commun. ACM 60, 6 (June 2017), 84–90. DOI:https://doi.org/10.1145/3065386

[18] Abate, Andrea F., et al. "2D and 3D face recognition: A survey." Pattern recognition letters 28.14 (2007): 1885-1906.

[19] Craven, Mark W., and Jude W. Shavlik. "Using neural networks for data mining." Future generation computer systems 13.2-3 (1997): 211-229.

[20] Dada, Emmanuel Gbenga, et al. "Machine learning for email spam filtering: review, approaches and open research problems." Heliyon 5.6 (2019): e01802.

[21] W. Shi et al., "Real-Time Single Image and Video Super-Resolution Using an Efficient Sub-Pixel Convolutional Neural Network," 2016 IEEE Conference on Computer Vision and Pattern Recognition (CVPR), Las Vegas, NV, USA, 2016, pp. 1874-1883. doi: 10.1109/CVPR.2016.207

[22] Goodfellow, Ian; Pouget-Abadie, Jean; Mirza, Mehdi; Xu, Bing; Warde-Farley, David; Ozair, Sherjil; Courville, Aaron; Bengio, Yoshua (2014). Generative Adversarial Networks (PDF). Proceedings of the International Conference on Neural Information Processing Systems (NIPS 2014). pp. 2672–2680.

[23] Ho, Jonathon; Ermon, Stefano (2016). "Generative Adversarial Imitation Learning". Advances in Neural Information Processing Systems: 4565–4573. arXiv:1606.03476. Bibcode:2016arXiv160603476H.

[24] Salimans, Tim; Goodfellow, Ian; Zaremba, Wojciech; Cheung, Vicki; Radford, Alec; Chen, Xi (2016). "Improved Techniques for Training GANs". arXiv:1606.03498

[25] Isola, Phillip; Zhu, Jun-Yan; Zhou, Tinghui; Efros, Alexei (2017). "Image-to-Image Translation with Conditional Adversarial Nets". Computer Vision and Pattern Recognition.

[26] Isola, Phillip, et al. "Image-to-image translation with conditional adversarial networks. CoRR abs/1611.07004 (2016)." arXiv preprint arXiv:1611.07004 (2016).

[27] Zhu, Jun-Yan, et al. "Unpaired image-to-image translation using cycle-consistent adversarial networks." Proceedings of the IEEE international conference on computer vision. 2017.

[28] Zhang, Xiwei, et al. "Exudate Detection in Color Retinal Images for Mass Screening of Diabetic Retinopathy." Medical Image Analysis, vol. 18, no. 7, 2014, pp. 1026–1043.

[29] Martinez-Perez, M.Elena, and Arturo Espinosa-Romero. "Three-Dimensional Reconstruction of Blood Vessels Extracted from Retinal Fundus Images." Optics Express, vol. 20, no. 10, 2012, pp. 11451–11465.

[30] Sierra, E., et al. "Dust Particle Artifact Detection and Removal in Retinal Images." Optica Pura y Aplicada, vol. 50, no. 4, 2017, pp. 379–387.

[31] Grinsven, Mark J. J. P. van, et al. "Fast Convolutional Neural Network Training Using Selective Data Sampling: Application to Hemorrhage Detection in Color Fundus Images." IEEE Transactions on Medical Imaging, vol. 35, no. 5, 2016, pp. 1273–1284.

[32] Yoo, Tae Keun, et al. "CycleGAN-Based Deep Learning Technique for Artifact Reduction in Fundus Photography." Graefes Archive for Clinical and Experimental Ophthalmology, vol. 258, no. 8, 2020, pp. 1631–1637.

[33] Sonali, et al. "An Approach for de-Noising and Contrast Enhancement of Retinal Fundus Image Using CLAHE." Optics and Laser Technology, vol. 110, 2019, pp. 87–98.

[34] Yoo, TaeKeun (2020), "A CycleGAN deep learning technique for artifact reduction in fundus photography", Mendeley Data, V1, doi: 10.17632/dh2x8v6nf8.1
http://dx.doi.org/10.17632/dh2x8v6nf8.1

[35] Salimans, Tim, and Diederik P. Kingma. "Weight normalization: A simple reparameterization to accelerate training of deep neural networks." arXiv preprint arXiv:1602.07868 (2016).

[36] Wu, Yuxin, and Kaiming He. "Group normalization." Proceedings of the European conference on computer vision (ECCV). 2018.

[37] Ba, Jimmy Lei, Jamie Ryan Kiros, and Geoffrey E. Hinton. "Layer normalization." arXiv preprint arXiv:1607.06450 (2016).

[38] Santurkar, Shibani, et al. "How does batch normalization help optimization?." arXiv preprint arXiv:1805.11604 (2018).

[39] Ioffe, Sergey, and Christian Szegedy. "Batch normalization: Accelerating deep network training by reducing internal covariate shift." International conference on machine learning. PMLR, 2015.

[40] Ulyanov, Dmitry, Andrea Vedaldi, and Victor Lempitsky. "Instance normalization: The missing ingredient for fast stylization." arXiv preprint arXiv:1607.08022 (2016).

[41] Huang, Xun, and Serge Belongie. "Arbitrary style transfer in real-time with adaptive instance normalization." Proceedings of the IEEE International Conference on Computer Vision. 2017.

[42] Zhang, Xiaohu, Yuexian Zou, and Wei Shi. "Dilated convolution neural network with LeakyReLU for environmental sound classification." 2017 22nd International Conference on Digital Signal Processing (DSP). IEEE, 2017.

[43] Maas, Andrew L., Awni Y. Hannun, and Andrew Y. Ng. "Rectifier nonlinearities improve neural network acoustic models." Proc. icml. Vol. 30. No. 1. 2013.

[44] Kingma, Diederik P., and Jimmy Ba. "Adam: A method for stochastic optimization." arXiv preprint arXiv:1412.6980 (2014).

[45] Ruder, Sebastian. "An overview of gradient descent optimization algorithms." arXiv preprint arXiv:1609.04747 (2016).

[46] He, Kaiming; Zhang, Xiangyu; Ren, Shaoqing; Sun, Jian (2015-12-10). "Deep Residual Learning for Image Recognition". arXiv:1512.03385 [cs.CV].

[47] He, Kaiming; Zhang, Xiangyu; Ren, Shaoqing; Sun, Jian (2016). "Deep Residual Learning for Image Recognition" (PDF). Proc. Computer Vision and Pattern Recognition (CVPR), IEEE. Retrieved 2020-04-23.

[48] Blum, Avrim L., and Pat Langley. "Selection of relevant features and examples in machine learning." Artificial intelligence 97.1-2 (1997): 245-271.

[49] Langley, Pat. "Selection of relevant features in machine learning." Proceedings of the AAAI Fall symposium on relevance. Vol. 184. 1994.

[50] Z. Wang, A.C. Bovik, H.R. Sheikh and E.P. Simoncelli, "Image quality assessment: from error visibility to structural similarity," IEEE Transactions on Image Processing, vol.13, no.4, pp. 600- 612, April 2004.

[51] H.R. Sheikh, M.F. Sabir and A.C. Bovik, "A statistical evaluation of recent full reference image quality assessment algorithms", IEEE Transactions on Image Processing, vol. 15, no. 11, pp. 3440-3451, Nov. 2006.

[52] Mittal, Anish, Anush K. Moorthy, and Alan C. Bovik. "Blind/referenceless image spatial quality evaluator." 2011 conference record of the forty fifth asilomar conference on signals, systems and computers (ASILOMAR). IEEE, 2011.

[53] Hochreiter, Sepp, et al. "Gradient flow in recurrent nets: the difficulty of learning long-term dependencies." (2001).